\documentclass[showpacs,preprintnumbers,amsmath,amssymb,floatfix]{revtex4-2}

\usepackage{graphicx}
\usepackage{dcolumn}
\usepackage{bm}
\usepackage{amssymb}
\usepackage{epsfig}
\usepackage{color}

\newcommand{\ba}{\begin{eqnarray}}
\newcommand{\ea}{\end{eqnarray}}
\newcommand{\be}{\begin{equation}}
\newcommand{\ee}{\end{equation}}
\newcommand{\bdisplay}{\begin{displaymath}}
\newcommand{\edisplay}{\end{displaymath}}

\newcommand{\eq}[1]{Eq.\,(\ref{#1})}
\newcommand{\fig}[1]{Fig.\,\ref{#1}}

\newcommand{\pbar}{\bar{p}}

\makeatletter
\def\eqnarray{\stepcounter{equation}\let\@currentlabel=\theequation
\global\@eqnswtrue
\tabskip\@centering\let\\=\@eqncr
$$\halign to \displaywidth\bgroup\hfil\global\@eqcnt\z@
  $\displaystyle\tabskip\z@{##}$&\global\@eqcnt\@ne
  \hfil$\displaystyle{{}##{}}$\hfil
  &\global\@eqcnt\tw@ $\displaystyle{##}$\hfil
  \tabskip\@centering&\llap{##}\tabskip\z@\cr}

\def\endeqnarray{\@@eqncr\egroup
      \global\advance\c@equation\m@ne$$\global\@ignoretrue}

\def\@yeqncr{\@ifnextchar [{\@xeqncr}{\@xeqncr[5pt]}}
\makeatother

\begin{document}

\title{Some remarks on Coulombic effects in $pp$ and $\pbar p$ scattering and the determination of $\rho$}

\author{ Loyal Durand}
\email{ldurand@hep.wisc.edu}
\altaffiliation{Present address: 415 Pearl Court, Aspen, CO 81611}
\affiliation{Department of Physics
University of Wisconsin-Madison, Madison, WI 53706}

\author{Phuoc Ha}
\email{pdha@towson.edu}
\affiliation{Department of Physics, Astronomy, and Geosciences, Towson University, Towson, MD 21252}

\begin{abstract}

We point out a very simple method for calculating the mixed Coulomb-nuclear corrections to the $pp$ and $\pbar p$ scattering amplitudes that has been missed in the extensive past work on this problem. The method expresses the correction in terms of a rapidly convergent  integral involving the inverse Fourier-Bessel transform of the nuclear amplitude and a known factor containing the Coulomb phase shift with form-factor corrections. The transform can be calculated analytically for the exponential-type model nuclear amplitudes commonly used in fits to the high-energy data at small momentum transfers, and gives very accurate results for the corrections. We examine the possible effects of the Martin zero in the real part of the nuclear amplitude, and the accuracy of the Bethe-West-Yennie phase approximation for the Coulomb-nuclear corrections. We then apply the method to a redetermination of the ratio $\rho$ of the real to the imaginary parts of the forward scattering amplitude in fits to high-energy ISR data previously analyzed using an approximate version of the correction. The only significant  changes relative the accuracy of those fits are at 52.8 GeV.  Our method is applicable more generally, and can be used also at lower energies and for proton-nucleus scattering.

\end{abstract}


\maketitle

%%%%%%%%%%%%%%%%%%%%%%%%%
%%%%%%%%%%%%%%%%%%%%%%%%%

\section{Introduction \label{sec:intro}} 

The effect of the Coulomb interaction in high-energy proton-proton and antiproton-proton  scattering has been studied by many authors over more than fifty years; see \cite{BetheCoulomb,Islam,WestYennie,Buttimore,Cahn,KL-Coulomb,KopeliovichTarasov,Petrov1,DH_CoulNucl,Nekrasov} and the many further references therein. A primary objective has been the use of Coulomb-nuclear interference effects to determine the ratios $\rho=\Re f_N/\Im f_N$ of the real to imaginary parts of the $pp$ and $\pbar p$ nuclear scattering amplitudes in the forward direction. The only other direct information on the real parts of the amplitudes is that obtained at much larger angles very near the observed dips in the differential cross sections. These dips are associated with diffraction zeros in the imaginary parts of the amplitudes where the scattering is dominated by the real part \cite{Ha_Re_f}. 

The most commonly used method for calculating the Coulomb-nuclear effects in data analyses appears at present to be that of Cahn \cite{Cahn}  as later modified by  Kundr\'{a}t and M. Lokaji\v{c}ek  \cite{KL-Coulomb}, but some earlier analyses, as of the data of Amos {\em et al.} \cite{Amos_ISR} from the CERN Intersecting Storage Rings (ISR), use an approximate version introduced  by Bethe \cite{BetheCoulomb} and later derived by West and Yennie \cite{WestYennie} through a diagrammatic analysis  in QED. The Cahn-Kundr\'{a}t-Lokaji\v{c}ek method is based on the use of the Fourier-Bessel convolution theorem to calculate the corrections that involve Coulomb and nuclear interactions simultaneously, and to include the effects of the nucleon charge form factors. The results, which involve delicate manipulations in their derivation to avoid singularities associated with the infinite range of the Coulomb interaction \cite{Cahn,Nekrasov}, and further complications in the subsequent evaluation of the convolutions, are not transparent; see, {\em e.g.} \cite{KL-Coulomb}, Eq.~(26) or \cite{TOTEM2019_Rho_13}, Eq.~(17).

We show here that the the full $pp$ and $\pbar p$ scattering amplitudes can each be written as the sum of a Coulomb and form-factor related term, the pure strong-interaction or ``nuclear'' amplitude, and a  mixed Coulomb-nuclear correction term, with
\be
\label{f_final1}
f(s,q^2)=-\frac{2\eta}{q^2}F_Q^2(q^2)+f_N(s,q^2)+\int_0^\infty db b J_0(qb)\left(e^{2i\delta_C'(b,s)+2i\delta_{FF}(b,s)}-1\right)\widehat{f}_N(b,s)+O(\eta^2).
\ee
Here $\eta=z_1z_2\alpha/v\rightarrow\alpha(-\alpha)$ for $pp$ ($\pbar p$) scattering at high energies, where $v=2pW/(W^2-2m^2)$. The Coulomb and form-factor associated phases $\delta_C'$ and $\delta_{FF}$ in the first factor in the integral are known, while ${\widehat f_N}(b,s)$ is the inverse Fourier-Bessel transform of the nuclear scattering amplitude.

The integral defining ${\widehat f}(b,s)$, \eq{inv_Bessel} below and the final integral in \eq{f_final1} are both rapidly convergent for realistic models of the nuclear amplitude $f_N(s,q^2)$, and ${\widehat f}(b,s)$ can be evaluated analytically for the exponential-type models  typically used to fit the observed cross sections at very high energies and small momentum transfers.  This approach to the calculation of the corrections is essentially obvious once it is recognized, but it has not been used in previous work including that of the present authors except in the context of a full eikonal model for the scattering \cite{Ha_Re_f}.   As we show by example, the correction term is small and easily calculated, significantly simplifying the analysis of Coulomb-nuclear interference in high-energy scattering relative to the methods now in use. Although our emphasis here is on the multi-GeV high-energy regime, the method can be used also at lower energies and for proton-nucleus scattering, the original objective in Bethe's work in \cite{BetheCoulomb}.

The layout of the paper is as follows. We first present the theoretical background of our method in Sec.~\ref{sec:theory}, then consider exponential-type models for the scattering amplitude in Sec.~\ref{subsec:exp_models} and check their accuracy in the calculation of the mixed Coulomb-nuclear corrections at small $q^2$ in Sec.~\ref{subsec:accuracy}. With that established, we use the models in Sec.~\ref{subsec:Martin} to investigate the sensitivity of the differential cross sections to Coulomb-nuclear interference, and the possible influence of the Martin zero in the real part the amplitude on the determination of the $\rho$ parameter. In Sec.~\ref{subsec:Bethe_phase}, we investigate the accuracy of the West-Yennie approximation for the Coulomb-nuclear correction based on the use of Gaussian form factors in a diagrammatic analysis \cite{WestYennie}, obtain the correct form-factor phase for this approximation, and show that the corrected West-Yennie result is essentially indistinguishable in the interference region from that obtained using the standard proton form factors. We apply the results to a reanalysis of the ISR data of Amos {\em et al.} \cite{Amos_ISR} which was based on the West-Yennie approximation in Sec.~\ref{subsec:fits}. The only significant changes are at 52.8 GeV where $\sigma_{\rm tot}$, $B$, and $\rho$ all change by amounts outside of the quoted uncertaines.

%%%%%%%%%%%%%%%%%%%%%%%%%
%%%%%%%%%%%%%%%%%%%%%%%%%

\section{Theoretical background \label{sec:theory}}

In the absence of significant spin effects, generally thought to be very small at high energies, the spin-averaged differential cross section  for proton-proton scattering can be written in terms of a single  spin-independent amplitude
\be
\label{f^tot}
f(s,q^2) = i\int_0^\infty db\,b\left(1-e^{2i\delta_{tot}(b,s)}\right)J_0(qb).
\ee
The total phase shift $\delta_{tot}$ is the sum of terms $\delta_C$ for pure Coulomb scattering, $\delta_{FF}$ for the effects of the charge form factors of the proton, and $\delta_N$ for the strong-interaction or nuclear scattering,
\be
\label{delta_total}
\delta_{tot}(b,s) = \delta_C(b,s)+\delta_{FF}(b,s) + \delta_N(b,s).
\ee
Here
\be
\label{delta_C} 
\delta_C(b,s) = \eta(\ln{pb}+\gamma) 
\ee
where $\gamma=0.5772\ldots$ is Euler's constant,  $\eta=z_1z_2\alpha/v\rightarrow \alpha$ $(-\alpha)$ for high-energy $pp$ $(\pbar p)$  scattering, and \cite{corr1} 
\be
\label{delta_FF}
\delta_{FF}(b,s) = \sum_{m=0}^3\frac{\eta}{2^m\Gamma(m+1)}(\mu b)^mK_m(\mu b)
\ee
for the standard proton charge form factor 
\be
\label{form_factor}
F_Q(q^2) = \frac{\mu^4}{\left(q^2+\mu^2\right)^2}
\ee
with $\mu^2=0.71$ GeV$^2$.

With our normalization, the differential scattering cross section is
\be
\label{dsigma/dq2}
\frac{d\sigma}{dq^2} = \pi\lvert f(s,q^2)\rvert^2,
\ee
where $q^2=-t$ is the square of the invariant momentum transfer and $W=\sqrt{s}$ is the total energy in the center-of-mass system.

The Coulomb amplitude corresponding to the phase shift $\delta_C(b,s)$ in \eq{delta_C} is $-(2\eta/q^2)e^{4i\eta\ln{(p^2/q^2})}$ \cite{DH_CoulNucl}.  The momentum-dependent factor $e^{4i\eta\ln{(p^2/q^2})}$  can be extracted from the complete scattering amplitude without affecting the  differential cross section, and the remaining amplitude written in the form  \cite{Ha_Re_f}
  \be
  \label{f_defined}
  f (s,q^2) = f'_C(s,q^2)+f_{FF}(s,q^2)+f_N(s,q^2)+f_N^{\,\rm Corr}(s,q^2),
  \ee
with $\delta_C\rightarrow \delta'_C$ now given in \eq{delta_total} by 
\be
\label{delta'_C}
\delta'_C(b,s) = \eta(\ln{(qb/2)}+\gamma).
\ee
With the overall phase $\left(4p^2/q^2\right)^{i\eta}$ which appears  in Eq.~(21) of \cite{DH_CoulNucl} removed, the Coulomb and form-factor terms combine as shown in \cite{DH_CoulNucl}, Sec.~IIC, to give
  \ba
   \label{f_Coul}
   f'_C(s,q^2)+f_{FF}(s,q^2) &=& -\frac{2\eta}{q^2}
  \left[1-\left(\frac{q^2}{q^2+\mu^2}\right)^{i\eta}\left(1-\frac{\mu^8}{(q^2+\mu^2)^4}\right)+O(\eta)\right] \\
  \label{f_Coul2}
  &=& -\frac{2\eta}{q^2}F_Q^2(q^2) +O(\eta^2,i\eta^2)
  \ea
 where the error terms are at most logarithmically divergent in $q^2$ for $q^2\rightarrow 0$.
 
 The purely nuclear amplitude, which is to be determined from fits to scattering data, is
  \be
   \label{f_N}
  f_N(s,q^2) = i\int_0^\infty db b \left(1-e^{2i\delta_N(b,s)}\right)J_0(qb).
  \ee
Finally,
  \be
 \label{corr_int}
  f_N^{\,\rm Corr}(s,q^2) = \int_0^\infty db b\left(e^{2i\delta'_C(b,s)+2i\delta_{FF}(b,s)}-1\right)\times i\left(1-e^{2i\delta_N(b,s)}\right)J_0(qb) 
 \ee
  is the mixed Coulomb-nuclear term.
  
 Our key observation is that the last factor in \eq{corr_int} is just the integrand for $f_N$ in \eq{f_N}, so may be evaluated as the inverse Fourier-Bessel transform $\widehat{f}_N(b,s) $ of the nuclear amplitude $f_N$ \cite{Bessel_inverse},
  \be
  \label{inv_Bessel}
\widehat{f}_N(b,s) = \int_0^\infty dq q f_N(s,q^2)J_0(qb) =  i\left(1-e^{2i\delta_N(b,s)}\right). 
 \ee
Thus,
  \ba
  \label{corr_int2}
    f_N^{\,\rm Corr}(s,q^2) &=& \int_0^\infty db b J_0(qb)\left(e^{2i\delta'_C(b,s)+2i\delta_{FF}(b,s)}-1\right)\widehat{f}_N(b,s) \\
    \label{corr_int3}
  &=& i\int_0^\infty db b\left(2\delta'_C(b,s)+2\delta_{FF}(b,s)\right)\widehat{f}_N(b,s)J_0(qb)+O(\eta^2).
\ea

 Both the integral in \eq{inv_Bessel} and the final integral in \eq{corr_int3} are expected to converge very rapidly for realistic models of the nuclear amplitude; no further manipulations are necessary to obtain a useful result. These simple results have been missed in previous work \cite{CKLapproach}, leading to unnecessary complications.

The function $\widehat{f}_N(b,s)$ can be determined for any successful phenomenological model for $f_N(s,q^2)$ by performing the inverse transform in \eq{inv_Bessel}.  This can be calculated analytically for the exponential-type models in $q^2$ commonly used in fitting the $pp$ and $\bar{p}p$ data at high energies and small momentum transfers, and some other models as well,   giving simple expressions that make the calculation of the Coulomb-nuclear correction straightforward by numerical evaluation of the remaining rapidly-convergent integral.  We will consider some examples in Sec.~III.

The mixed Coulomb-nuclear correction can also be evaluated efficiently numerically for models in which $\widehat{f}_N(b,s)$ cannot be calculated analytically. In that case,
\be
\label{double_int}
  f_N^{\,\rm Corr}(s,q^2) = \int_0^\infty db b\left[ \left(e^{2i\delta'_C(b,s)+2i\delta_{FF}(b,s)}-1\right)\times \int_0^\infty dq' q' f_N(s,q'^2)J_0(q'b)\right]J_0(qb),
\ee
 where it is essential that the inner integral over $q'$ in \eq{double_int} be evaluated first. This integral converges rapidly for any reasonable model for $f_N$ that describes the rapid, nearly exponential, fall of the differential cross sections with increasing $q^2$ observed at high energies, and gives a result that  vanishes rapidly for large $b$ as  expected from the long-range behavior of strong interactions. The second integral over $b$ is therefore also expected to converge rapidly.  This will be seen explicitly in the examples in Sec.~III.
 
 The order of the integrations is crucial: the Coulomb plus form factor term in parentheses in \eq{double_int} does not provide convergence at large $b$ if one tries to integrate in the opposite order, and one encounters the singularities that caused trouble in Cahn's approach  and its later modifications \cite{Cahn,CKLapproach}. The double integral in \eq{double_int} converges well when performed in the order specified, and can easily be evaluated numerically. 
 
 With this approach, the corrected full amplitude $f(s,q^2)$ can be determined very simply for a model $f_N(s,q^2)$, and $\rho$ then determined through  Coulomb-nuclear interference in subsequent fits to $d\sigma/dq^2$.

 %%%%%%%%%%%%%%%%%%%%%%%%%
%%%%%%%%%%%%%%%%%%%%%%%%%

\section{Simple calculation of the mixed Coulomb-nuclear corrections \label{sec:examples}}

%%%%%%%%%%%%%%%%%%%%%%
%%%%%%%%%%%%%%%%%%%%%%

\subsection{Exponential-type models for $f_N$ \label{subsec:exp_models}}

Consider as an example of a trial nuclear amplitude $f_N$  the simple exponential model
\be
\label{exp_model}
f_N^{\,\rm exp}(s,q^2)=(i+\rho)\sqrt{A/\pi}e^{-\frac{1}{2}Bq^2}, \quad d\sigma/dq^2=A\left(1+\rho^2\right)e^{-Bq^2},
\ee
with $A,\,B$ and $\rho$ functions of $s$ but independent of $b$.  This model has been used over very wide range of energies to fit experimental data on the $pp$ and $\pbar p$ differential cross sections at small $q^2$ to determine the forward slope parameters $B=-d(\ln{d\sigma/dq^2})/dq^2\vert_{q^2=0}$, the total cross sections $\sigma_{\rm tot}= 4\pi\Im f_N(s,0)=4\pi\sqrt{A/\pi}$, and to determine the ratios $\rho(s)=\Re f_N(s,0)/\Im f_N(s,0)$ of the real to the imaginary parts of the forward amplitudes from Coulomb-nuclear interference effects. See, for example, \cite{Amos_ISR} and \cite{TOTEM2016} for examples at 52.8 GeV and 8 TeV.

For this model the inverse Fourier-Bessel transform in \eq{inv_Bessel} is
\be
\label{exp_model2}
\widehat{f}_N^{\,\rm exp}(b,s) = (i+\rho)\int_0^\infty dq q \sqrt{\frac{A}{\pi}}e^{-\frac{1}{2}Bq^2}J_0(qb)=(i+\rho)\sqrt{\frac{A}{\pi}}\frac{1}{B}e^{-b^2/2B}.
 \ee
The remaining integral over $b$ in \eq{corr_int2} converges exponentially and is easily evaluated numerically  to get the mixed Coulomb-nuclear correction term  $f_N^{\,\rm Corr}(s,q^2)$. 

 The simple exponential model can be extended to 
 \be
 \label{fN_curvature}
 f_N^{\,exp''}(b,q^2)= (i+\rho)\sqrt{A/\pi}e^{-\frac{1}{2}B'q^2}(1+\frac{1}{2}Cq^4-\frac{1}{2}Dq^6+\cdots)
 \ee
 to include the curvature corrections introduced in \cite{TOTEM2015} and derived theoretically in \cite{BDHHCurvature}. This is necessary to obtain a precise fit to $d\sigma/dq^2$ away from very small momentum transfers. This model, an expanded form of the result in \cite{BDHHCurvature}, can again be handled analytically using the formula (\cite{Watson}\,13.3(3))
 \be
 \label{watson_formula}
 \int_0^\infty J_0(at)e^{-p^2t^2}t^{2n+1}dt=\frac{\Gamma(n+1)}{2(p^2)^{n+1}}e^{-a^2/4p^2}\ _1F_1(-n;1;a^2/4p^2),
 \ee
 where the hypergeometric series terminates after the $n$-plus-first term, $n=0,\,1,\ldots$. While the curvature corrections can affect the overall fit to  data, especially the values of $B$ and $\sigma_{tot}$, and should be included in $f_N(s,q^2)$ to obtain precise fits to the differential cross section away from the forward dir3ction, they are too small to affect $f_N^{\rm Corr}$ significantly and can be ignored in the calculation of that term.
  
The inner integral in \eq{double_int} can also be evaluated analytically for some other models, for example,  models that use the functions  ${\rm exp}(-\sqrt{b^2/a^2+c^2})/ \sqrt{b^2/a^2+c^2}$ introduced by Ferreira, Kohara, and Kodama \cite{Ferreira}.  However, for most models, such as those based on Regge theory or comprehensive eikonal fits to the data, the integrals must be evaluated numerically. As shown by the simple exponential model, the inner and outer integrals may still be expected to converge very rapidly as functions of $q'$ and $b$ for realistic $f_N$.

%%%%%%%%%%%%%%%%%%%%%%
%%%%%%%%%%%%%%%%%%%%%%

\subsection{Accuracy of the Coulomb-nuclear corrections for exponential models \label{subsec:accuracy}}

%%%%%%%%%%%%%%%%%%%%%%

 The exact eikonal expression for the Coulomb-nuclear correction $f_N^{\,\rm Corr}(s,q^2)$ to the $pp$ scattering amplitude is given in \eq{corr_int}. 
The approximate expression obtained using the the simple exponential model is 
 \be
 \label{fCorr_exp}
 f_N^{\rm Corr}(s,q^2)=2i\eta(i+\rho)\sqrt{\frac{A}{\pi}}\frac{1}{B}\int_0^\infty db b\left[\ln{\frac{qb}{2}}+\gamma+\sum_{m=0}^3\frac{1}{2^m\Gamma(m+1)}(\mu b)^mK_m(\mu b)\right]e^{-b^2/2B}J_0(qb)+O(\eta^2).
 \ee
 The rapid convergence of the integral in the latter associated with the Gaussian factor is clear. The convergence is further enhanced by the exponential decay of the hyperbolic Bessel functions $K_m(\mu b)$ with increasing argument, so the numerical evaluation of the integral is straightforward \cite{analytic_evaluation}. 
 
 To test the accuracy of the expression in \eq{fCorr_exp},  we have compared the results for $f_N^{\rm Corr}$ obtained using the eikonal model of Block {\em et al.} \cite{eikonal2015,eikonal_update}  to those obtained in the exponential model  with the parameters obtained by fitting the  differential cross sections  $d\sigma/dq^2$ in eikonal model as in a fit to the data. The full eikonal model satisfies the constraints of unitarity, analyticity, and crossing symmetry, fits the data on $\sigma_{\rm tot}$, $\sigma_{\rm elas}$, $B$, and $\rho$ for $pp$ and $\pbar p$ scattering from 5 GeV to 57 TeV, and gives a good description of the differential scattering cross sections and dip structure even though the data on $d\sigma/dq^2$ other than B were not used in the fit. It is taken here as representing the experimental data. 

\begin{figure}[htbp]
\includegraphics{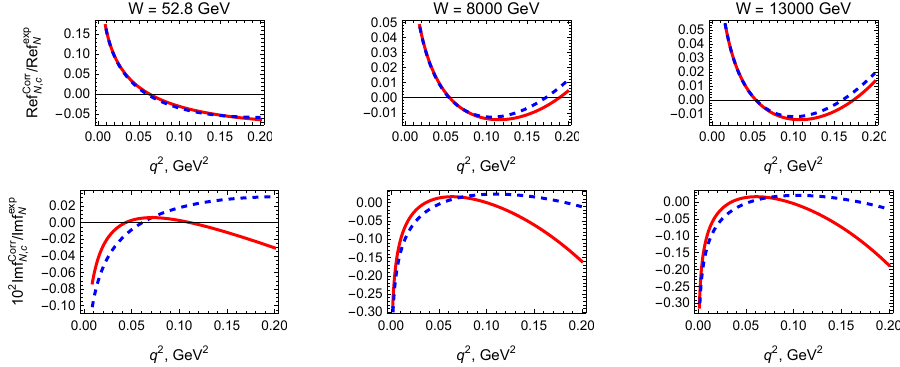}
 \caption{Top row: ratios of the real parts of the mixed Coulomb-nuclear corrections $f_N^{\,\rm Corr}(s,q^2)$  to the real parts of the $pp$ elastic scattering amplitude calculated using the eikonal model of Block {\em et al.} \cite{eikonal2015,eikonal_update} for the nuclear amplitude $f_N(s,q^2)$ (solid red lines) and the simple exponential model $f_N^{\rm exp}(s,q^2)$ of \eq{exp_model} (dashed blue line), to the real parts of $f_N^{\rm exp}(s,q^2)$. Bottom row: $10^2$ times the corresponding ratios of the imaginary parts of  $f_N^{\,\rm Corr}(s,q^2)$ calculated using the eikonal model  (solid red line) and the exponential model (dashed blue line) to the imaginary part of the exponential model.  }
 \label{fig1}
\end{figure}

In \fig{fig1} we show the ratios of the real and imaginary parts of the mixed Coulomb-nuclear corrections $f_N^{\,\rm Corr}(s,q^2)$ calculated using the eikonal model of Block {\em et al.} \cite{eikonal2015,eikonal_update} (red solid curves) and the exponential model of \eq{exp_model} with the parameters taken from the fits (blue dashed curves), to the  real and imaginary parts of the simple exponential model  $f_N^{\,\rm exp}(s,q^2)$. The full eikonal model satisfies the constraints of unitarity, analyticity, and crossing symmetry, fits the data on $\sigma_{\rm tot}$, $\sigma_{\rm elas}$, $B$, and $\rho$ for $pp$ and $\pbar p$ scattering from 5 GeV to 57 TeV, and gives a good description of the differential scattering cross sections and dip structure even though the data on $d\sigma/dq^2$ other than B were not used in the fit. It is taken here as representing the experimental data.

%%%%%%%%%%% FIG. 1 - magnitudes %%%%%%%

%%%%%%%%%%%%%%%%%%%%%

 As seen in the top row in \fig{fig1}, the real parts of the corrections calculated using the simple exponential model and \eq{corr_int2} agree remarkably well at small $q^2$ with those calculated in the eikonal model using the expression in \eq{corr_int} with the eikonal phase shift. This agreement  would be expected. The real part of the correction is associated mainly with the imaginary part of the nuclear scattering amplitude as may be seen by expanding the exponential in the factor in parentheses in \eq{corr_int2} to first order in the small quantity $\eta$. Since $\Im f_N\gg\lvert\Re f_N\rvert$, the very good fit of the exponential model to $d\sigma/dq^2$ over the range of small $q^2$ over which the cross section decreases rapidly \cite{Amos_ISR,TOTEM2016}   implies a correspondingly good fit  to $\Im f_N$ over that region, hence an accurate result for the real part of the correction term.  
 
 This is important.  The Coulomb plus form-factor amplitude $f'_C(s,q^2)$ in \eq{f_Coul} is real up to terms of order $\eta^2$, so the Coulomb-nuclear interference involves only $\Re f_N^{Corr}(s,q^2)$ in practice.
 
\begin{figure}[th]
\centering
\includegraphics[width=8cm]{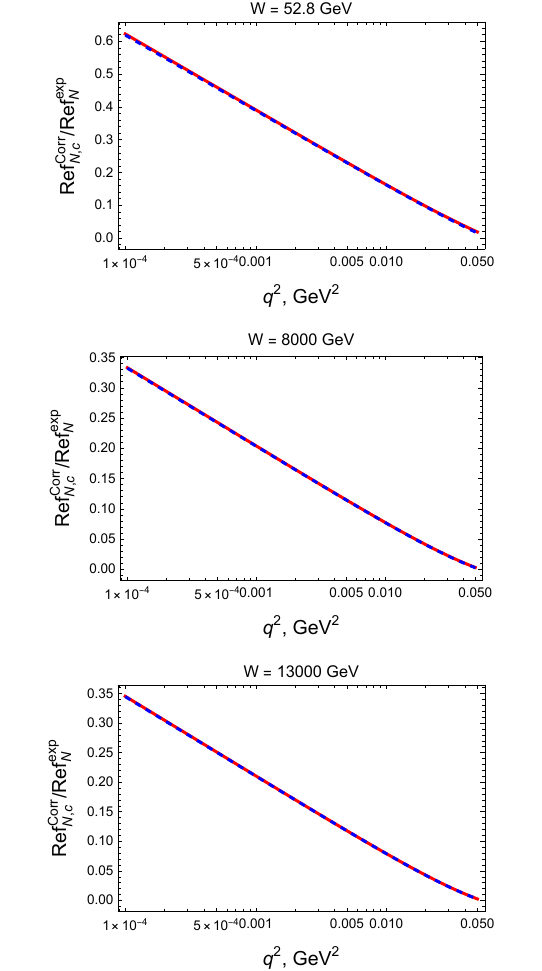}
\includegraphics[width=8cm]{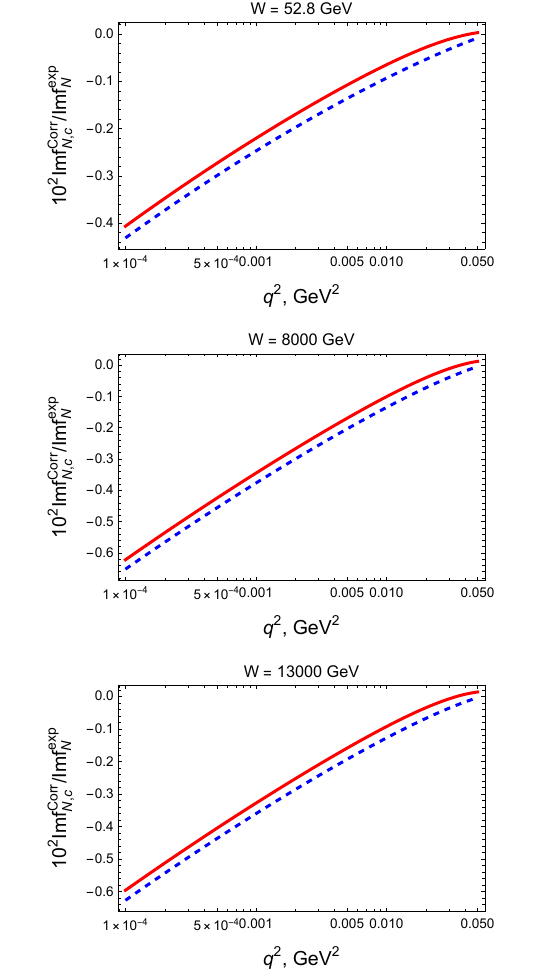}
\caption{Left-hand column: ratios of the real parts of the mixed Coulomb-nuclear corrections $f_N^{\,\rm Corr}(s,q^2)$  to the $pp$ elastic scattering amplitude calculated using the eikonal model (solid red lines) and the simple exponential model  of \eq{exp_model} (dashed blue line), to the real parts of $f_N^{\rm exp}$ in the region of Coulomb-nuclear interference. Right-hand column:  $10^2$ times the ratios of the imaginary parts of  $f_N^{\,\rm Corr}(s,q^2)$ calculated using the eikonal model  (solid red line) and the exponential model (dashed blue line) to the imaginary part of the exponential model.  }
\label{fig2}
\end{figure}

 The imaginary parts of the mixed Coulomb-nuclear correction found using the exponential model are considerably less accurate, but are  quite small with errors of less than a part in $10^3$ at small $q^2$ as seen in the bottom row in \fig{fig1} and the right-hand column in \fig{fig2}. They arise  from the real part of $f_N$, small compared to the imaginary part, and give negligible corrections to the imaginary parts of the full amplitude. This is not involved in the Coulomb-nuclear interference, hence, in the determination of $\rho$. 
 
 The relative inaccuracy of the corrections to the imaginary part of the amplitude seen in \fig{fig1} results from the poor description of $\Re f_N$ given by the exponential model. As expected from a theorem of Martin \cite{Martin} and seen in the eikonal model, there is a diffraction zero in $\Re f_N$ between $q^2=0$ and the first diffraction zero in $\Im f_N$. This is not evident in the differential cross sections because of the small size of $\Re f_N$ relative to $\Im f_N$ for $q^2$ below the dip region, but still leads to a much more rapid decrease of $\Re f_N$ than $\Im f_N$ as $q^2$ increases from 0. 

The Martin zero is not included in $f_N^{\rm exp}$, \eq{exp_model}, and can only be incorporated by using information on the position of the zero that is not available from experiment. We have found that simply including a separate exponential term for the real part of the model amplitude with a magnitude $\rho$ relative to the imaginary part and a slope parameter $B_R$ matched to that in the eikonal model eliminates most of the errors in the comparisons in the bottom row in \fig{fig1}. Other potential effects of the zero will be explored more below.

In \fig{fig2} we show  ratios of the real and imaginary parts of the mixed Coulomb-nuclear corrections $f_N^{\,\rm Corr}(s,q^2)$ to the real and imaginary parts of $f_N^{\rm exp}$ for the eikonal and exponential models at very small $q^2$, the region of the observed Coulomb-nuclear interference. The agreement of the results for the real parts is excellent.  The corrections to the real part of the amplitude are substantial and diverge logarithmically at small $q^2$ because of the term $\ln{(qb/2)}$ in $\delta'_C$, \eq{delta'_C}. While small compared to the Coulomb term itself, the corrections in $\Re f_N^{\,\rm Corr}$ cannot be neglected in analyses of Coulomb-nuclear interference.

%%%%%%%% FIG. 2 %%%%%%%%%%%%%%
%%%%%%%%%%%%%%%%%%%%%%

%%%%%%%%%%%%%%%%%%%%%%
%%%%%%%%%%%%%%%%%%%%%%
%%%%%%%%%%%%%%%%%%%%%%

\section{Comments on Coulomb-nuclear interference \label{sec:sensitivity}}

%%%%%%%%%%%%%%%%%%%%%%

\subsection{Small $q^2$ and the Martin zero \label{subsec:Martin}}

\begin{figure}[htbp]
\centering
\includegraphics[width=8cm]{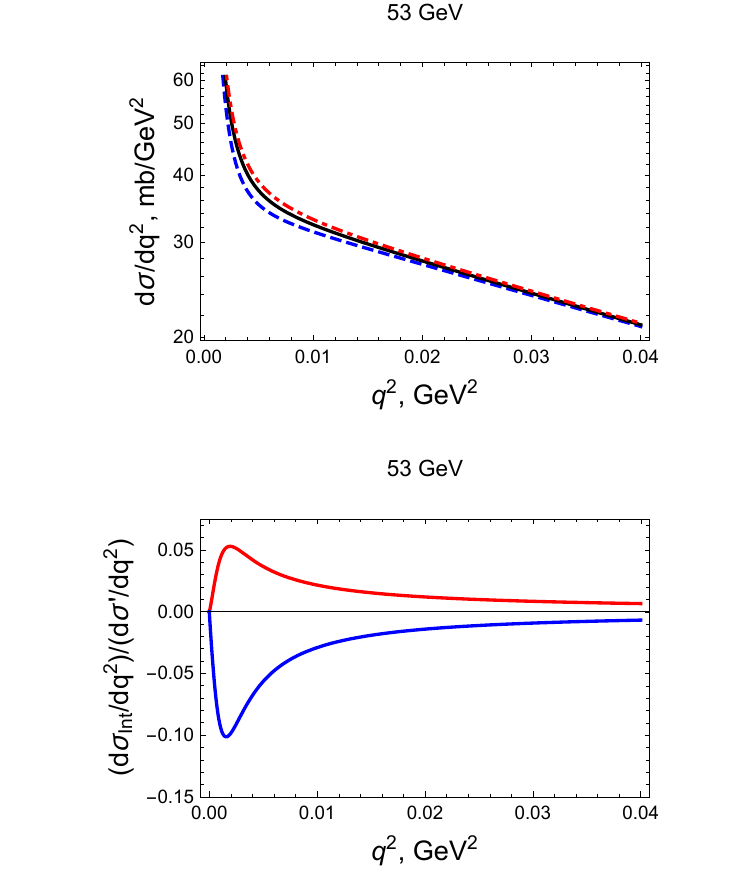}
\includegraphics[width=8cm]{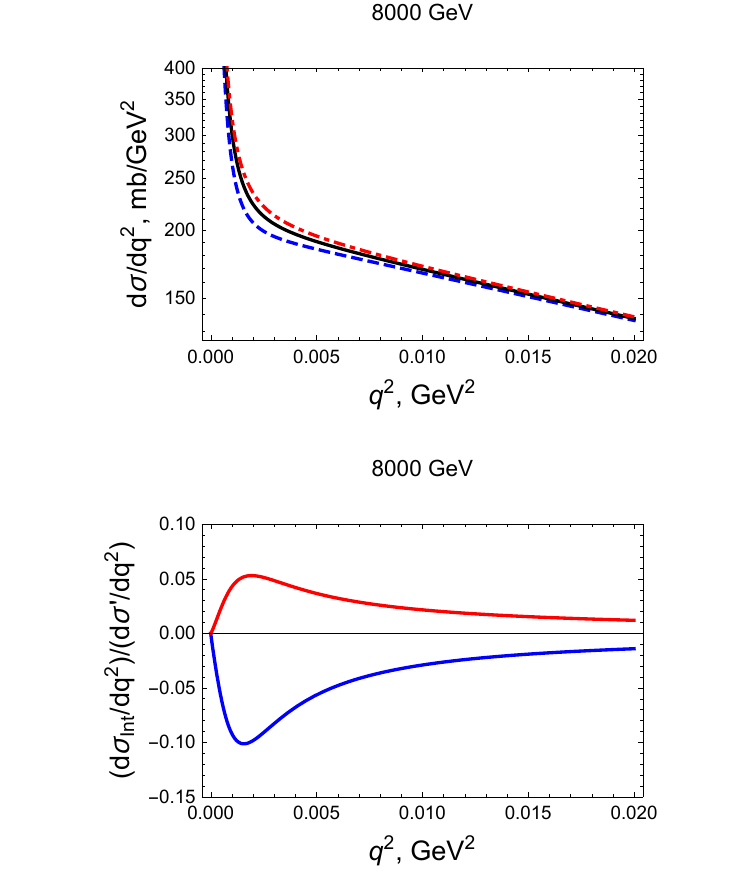}
\caption{Plots of the $pp$ and $\pbar p$ sensitivity ratios at 53 GeV (left-hand column) and 8 TeV (right-hand column). The top row shows  the $pp$ and $\pbar p$ differential  cross sections with Coulomb-nuclear interference included (dashed blue and dot-dashed red curves respectively), and the differential  cross section with the interference terms omitted (solid black curve). The bottom row shows the ratios of the interference terms to the differential  cross sections with those terms omitted for $\pbar p$ (top red curves) and $pp$ (bottom blue curves) scattering. The parameters used for $pp$ scattering were those determined in the fits in \cite{Ha_Re_f}; the same parameters were used for $\pbar p$ scattering to give a comparison of the sensitivites to Coulomb-nuclear interference. }
\label{Fig3}
\end{figure}

The region in which  Coulomb-nuclear interference effects in $d\sigma/dq^2$ are large enough to be detectable in the presence of experimental uncertainties is rather limited given the small size of the Coulomb amplitude. This is evident in the top panels in \fig{Fig3} where we show the  $pp$ and $\pbar p$ differential  cross sections  at 53 GeV and 8 TeV with and without the inclusion of the interference term. The $pp$ differential cross sections are calculated using the simple exponential model for $f_N$ with parameters from the fits in \cite{Ha_Re_f}; the same nuclear amplitudes are used for $\pbar p$ scattering to emphasize the different effects of the Coulomb-nuclear interference in the two cases. For reference, the statistical experimental uncertainties in the $pp$ differential  cross sections are less than 1 mb/GeV$^2$ (2 mb/GeV$^2$) at 53 GeV (8 TeV) over the ranges shown.

In the lower panels of \fig{Fig3} we show the ratio
\be
\label{interference_ratio}
(d\sigma_{int}/dq^2)\big/(d\sigma'/dq^2) = -\frac{4\pi\eta}{q^2}F_Q^2(q^2)\,\Re\Big[f_N(s,q^2)+f_N^{\rm Corr}(s,q^2)\Big]\Big/
(d\sigma'/dq^2)
\ee
of the interference term to the differential cross section $d\sigma'/dq^2$ with the interference term omitted. It is simple to show for the exponential model that this ratio  has a maximum at $q^2\approx 2\eta\sqrt{\pi/A}$, dropping off sharply for smaller $q^2$ and decreasing less rapidly for larger $q^2$ as seen in the figure.

%%%%%%%%%% FIG. 3 - sensitivity %%%%%%%
%%%%%%%%%%%%%%%%%%%%%%%%%%%%%%%%%%%%%%%%%%

It it clear that the regions of maximum sensitivity to the effects of the Coulomb-nuclear interference are at very low values of $q^2$. This has the effect of suppressing the potential effects of the Martin zero \cite{Martin} expected in the real part of the nuclear scattering amplitude.  Pacetti {\em et al.} \cite{Pacetti} have suggested that the neglect of the effects of this zero on the magnitude of the observed interference could account for  the low values of $\rho$ found in the analysis of the TOTEM experiments at 8 and 13 TeV. See also Kohara, Ferreira, and Rangel \cite{Kohara}. 

The location of the Martin zero and the energy and momentum-transfer dependence of $\rho(s,q^2)$ were examined in detail in \cite{DH_CoulNucl}, Sec.~III, where we gave useful parametrizations of both the location of the zero, and the shape of $\rho(s,q^2)$ obtained in the eikonal model of \cite{eikonal2015}.  Subsequent calculations showed no significant effects of the zero on the determination of $\rho$ in fits to the data, a result attributed to the very small values of $q^2$ at which the fits are most sensitive.

\begin{figure}[htbp] 
\includegraphics{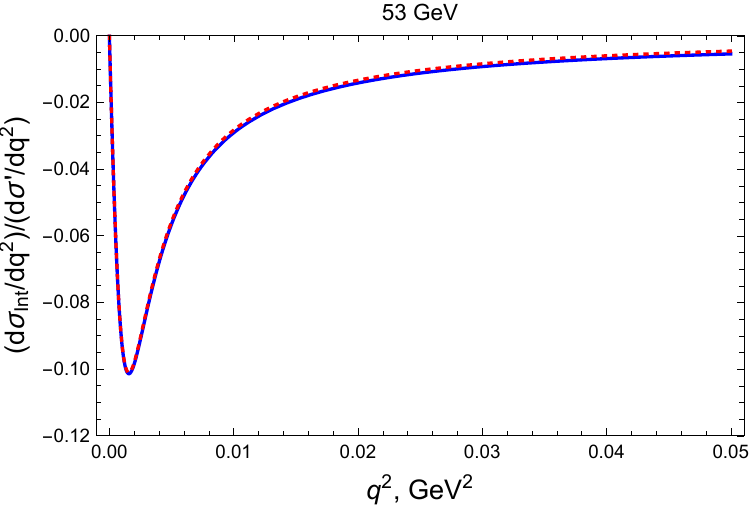}
 \caption{ Ratio of the the Coulomb-nuclear interference term in the differential $pp$ cross section  to that cross section with the interference term omitted, \eq{interference_ratio}, at 53 GeV.  Solid (blue) curve: ratio with $\rho$ independent of $q^2$. Dashed (red) curve; ratio with the Martin zero included with $\rho=(1-q^2/q_M^2)$ with $q_M^2=0.319$ GeV$^2$ the location of the Martin zero in the comprehensive eikonal model of Block {\em et al.} \cite{eikonal2015,eikonal_update}.}.
 \label{Martin_ratio}
\end{figure}

The effect of the Martin zero at $q^2=0.319$ GeV$^2$  on the ratio in \eq{interference_ratio} at 53 GeV is shown in \fig{Martin_ratio}; it is clearly negligible in the region of greatest sensitivity to the Coulomb-nuclear interference. The effects are  similarly small at  8 and at 13 TeV, where the regions of maximum sensitivity to the are shifted to still smaller $q^2$ as in \fig{Fig3}, staying well below the locations of the respective zeros at $q^2= 0.169 $ GeV$^2$ ($q^2=0.156 $ GeV$^2$). 

We conclude that the Martin zero and the associated rapid variation of $\rho(s,q^2)$ as a function of $q^2$ can safely be ignored in data analyses at small $q^2$.

%%%%%%%%%%% FIG. 4 - Martin ratio %%%%%%%
 
%%%%%%%%%%%%%%%%%%%%%

%%%%%%%%%%%%%%%%%%%%

\subsection{ The Bethe-West-Yennie phase and $f_N^{\rm Corr}(s,q^2)$
 \label{subsec:Bethe_phase}}

Following Bethe \cite{BetheCoulomb}, it has been common in discussions of Coulomb effects in $pp$ and $\pbar p$ scattering  to try to express the mixed Coulomb-nuclear corrections through a phase factor $e^{\pm i\Phi}$ multiplying either the Coulomb (+) or the nuclear amplitude (-) \cite{Phi_definition},  with, {\em e.g.},
\be
\label{Bethe_phase}
f(s,q^2)\approx -\frac{2\eta}{q^2}F_Q^2(q^2)+e^{-i\Phi} f_N(s,q^2).
\ee
Although the complete nuclear amplitude with the mixed Coulomb-nuclear correction included,
\be
\label{nucl_combined}
f_N^{\rm Comp}(s,q^2) = f_N(s,q^2)+f_N^{\rm Corr}(s,q^2)=\int_0^\infty db be^{2i\delta'(b,s)+2i\delta_{FF}(b,s)}{\widehat f}_N(b,s)J_0(qb),
\ee
does not factor in general because of the integration, the form of the last term in \eq{Bethe_phase} is suggested by the observation that $b{\widehat f}_Nb,s)$ is sharply peaked for $b\approx \sqrt{B}$, while  the Coulomb phase in the integrand  varies less rapidly and can reasonably be approximated by its value at the peak. This approximation, while suggestive, is less accurate for the phase associated with the form factors, \eq{corr_int2}, which varies significantly over the region in which $\widehat{f}_N(b,s)$ is large.  

Bethe  made his observations with respect to the Coulomb phase more precise using a WKB-type argument, and included an estimate of the effects of the form factors on the mixed Coulomb-nuclear effects using a Gaussian representation of those functions. The result was a phase for $pp$ scattering of the form $\Phi\approx -\eta\left(\ln{(q^2B/2)} +\gamma+{\rm constant}\right)$, with the constant not precisely determined. 

Given the lack of factorization and the smallness of the Coulomb and form-factor phases in \eq{corr_int2}, both proportional to $\eta$, it is useful to expand the exponential  to first order in $\eta$, the accuracy considered in \cite{BetheCoulomb}. This gives $f_N(s,q^2)-i\Phi f_N(s,q^2)]+\cdots$, with the second $O(\eta)$ term now to be regarded as Bethe's approximation for $f_N^{\,Corr}(s,q^2)$ \cite{factorization}.

In their diagrammatic analysis in perturbative QED, West and Yennie \cite{WestYennie} introduced the same Gaussian form factors as Bethe in the diagram which describes the Coulomb interaction between the nucleons, with the slopes of the Gaussians chosen to match the observed slopes of the proton form factor for $q^2\rightarrow 0$, leading to their expression for the effective Bethe phase and the correction term,

\be
\label{fCorr_WY}
f_{N,\,WY}^{\,\rm Corr}(s,q^2)\approx i\eta \left[\ln{\left(q^2\left(\frac{B}{2}+\frac{4}{\mu^2}\right)\right)}+\gamma\right] f_N(s,q^2).
\ee
No direct factorization of the complete amplitude in the form in \eq{Bethe_phase} was implied.

 As West and Yennie noted in \cite{WestYennie}, their treatment of the form factors was {\em ad hoc}; it was not really clear how the form-factor effects could be included consistently in a diagrammatic analysis. This is not the case in an eikonal treatment of the scattering as seen, for example, in a Glauber-type treatment \cite{Glauber} where the form factors contribute a separate eikonal phase as in \eq{f_final1}.

The eikonal phase $2\delta_{FF}$ \cite{corr1} for the standard proton form factors, \eq{delta_FF}, was determined in \cite{DH_CoulNucl}. We follow the same procedure here, and write the usual Born expression for the Coulomb interaction with Gaussian form factors as
\ba
\label{Gauss_phase1}
-\frac{2\eta}{q^2\,}e^{-q^2/\nu^2}&=& -\frac{2\eta}{q^2} -\frac{2\eta}{q^2} \left(e^{-q^2/\nu^2}-1\right) \\
\label{Gauss_phase2}
&=& -\frac{2\eta}{q^2} + \frac{2\eta}{\nu^2}\int_0^1dt e^{-q^2t/\nu^2},\qquad \nu^2=\mu^2/4,
\ea
where we have matched the slopes of the Gaussians and $F_Q^2(q^2)$, \eq{form_factor}, for $q^2\rightarrow 0$ to determine $\nu^2$. The two terms in these equations correspond to the Fourier-Bessel transforms of $2i\delta'_C$ and $2i\delta_{FF}$ as is evident from a first-order expansion of the phases in \eq{nucl_combined}.  The phases are given by the inverse transforms which can be calculated analytically for both the standard and Gaussian proton form factors. 

We note that the same decomposition of the product $-(2\eta/q^2)\left(F_1(q^2)F_2(q^2)\right)$ can be used to obtain the form-factor phase in the case of scattering of different particles, {\em e.g.}, $\pi p$ scattering as in West-Yennie \cite{WestYennie} or proton-nucleus scattering as in Bethe's original work \cite{BetheCoulomb}. It may be necessary in some cases to calculate the inverse Fourier-Bessel transforms numerically, but the integrals involved converge rapidly.

The pure Coulomb  term in \eq{Gauss_phase2} can be treated as in \cite{DH_CoulNucl} to obtain the Coulomb phase in \eq{delta_C}. To obtain $2\delta_{FF}$, we calculate the inverse Fourier-Bessel transform of the second term in \eq{Gauss_phase2}. The result, with a change of the integration variable to $u=1/t$, is
\be
\label{E1}
2\delta_{FF}^{\rm Gauss} =\eta\int_1^\infty \frac{du}{u}e^{-(\nu b^2/4)u} = \eta E_1\left(\frac{\nu^2b^2}{4}\right)
\ee
where $E_1(z)$ is the exponential integral function, \cite{dlmf} Sec.~6.2.

This  approximation for $2\delta_{FF}$ is good in the most relevant part of impact parameter space even though the matching condition was imposed on the Gaussian form factor as a  function of $q^2$ rather than $b$. Thus the expression in \eq{E1} differs from the phase  for the standard proton form factor in \eq{delta_FF} by less than a percent for $b=10^{-4}$ GeV$^{-1}$, and by $~6$\% for $b\approx 2$ GeV$^{-1}$ near the peak in the integrand in \eq{nucl_combined}, but cuts off more sharply at large $b$.

To obtain $f_{N,\rm Gauss}^{\rm Corr}$ to $O(\eta)$,  we multiple $2i\delta_{FF}$ by ${\widehat f}_N(b,s)$ and calculate the Fourier-Bessel transform. For the exponential model this gives
\be
\label{Gauss_corr1}
f_{N,FF}^{\rm Corr}(s,q^2) = -\frac{\eta}{2B} \sqrt{\frac{A}{\pi}}(1-i\rho)\int_1^\infty \frac{du}{u(u+\beta)}e^{-(q^2/\nu^2)/(u+\beta)} 
\ee
where $\beta=2/\nu^2B$. With the change of variable $u=\beta y/(1-y)$, this becomes
\ba
\label{Gauss_corr2}
f_{N,FF}^{\rm Corr}(s,q^2) &=& i\eta f_N(s,q^2)\int_{1/(1+\beta)}^1\frac{dy}{y}e^{(q^2B/2)y} \\
\label{Gauss_corr3}
&=& i\eta\left[\ln{\left(\frac{2}{B}\left(\frac{B}{2}+\frac{1}{\nu^2}\right)\right)}+\int_{1/(1+\beta)}^1\frac{dy}{y}\left(e^{(q^2B/2)y}-1\right)\right]f_N(s,q^2) \\
\label{Gauss_corr4}
&=& i\eta\left[\ln{\left(\frac{2}{B}\left(\frac{B}{2}+\frac{1}{\nu^2}\right)\right)}+\sum_{k=1}^\infty\frac{1}{k\,k!}\left(1-\left(\frac{1}{1+\beta}\right)^k\right)\left(\frac{q^2 B}{2}\right)^k\right]f_N(s,q^2).
\ea

When we add in the pure Coulomb contribution $i\eta(\log{(q^2B/2)}+\gamma)f_N(s,q^2)$, we obtain 
\be
\label{Gauss_corr6}
f_{N,\rm Gauss}^{\rm Corr}(s,q^2) =i\eta\left[\ln{\left(q^2\left(\frac{B}{2}+\frac{4}{\mu^2}\right)\right)}+\gamma+\sum_{k=1}^\infty\frac{1}{k\,k!}\left(1-\left(\frac{1}{1+\beta}\right)^k\right)\left(\frac{q^2 B}{2}\right)^k\right]f_N(s,q^2).
\ee
The real part of the first two terms gives the West-Yennie result for the Coulomb-nuclear correction \cite{WestYennie}. The remaining series gives non-negligible $q^2$-dependent corrections. The appearance of $f_N$ as an overall factor appears to be special to the exponential model for $f_N(s,q^2)$---exponential in $q^2$ or  Gaussian in $q=\sqrt{q^2}$\,---with Gaussian form factors.

We do not have an analytic expression for $f_N^{\rm Corr}$ for the standard form factor. Such an expression can be obtained in principle \cite{analytic_evaluation}, but is sufficiently complex that it is simpler to use the easily calculable form in \eq{corr_int3} with the phase $\delta_{FF}$ in \eq{delta_FF}.

%%%%%%%%%%% Fig. 5 West-Yennie phase   %%%%%%%

\begin{figure}[tbh]
\includegraphics{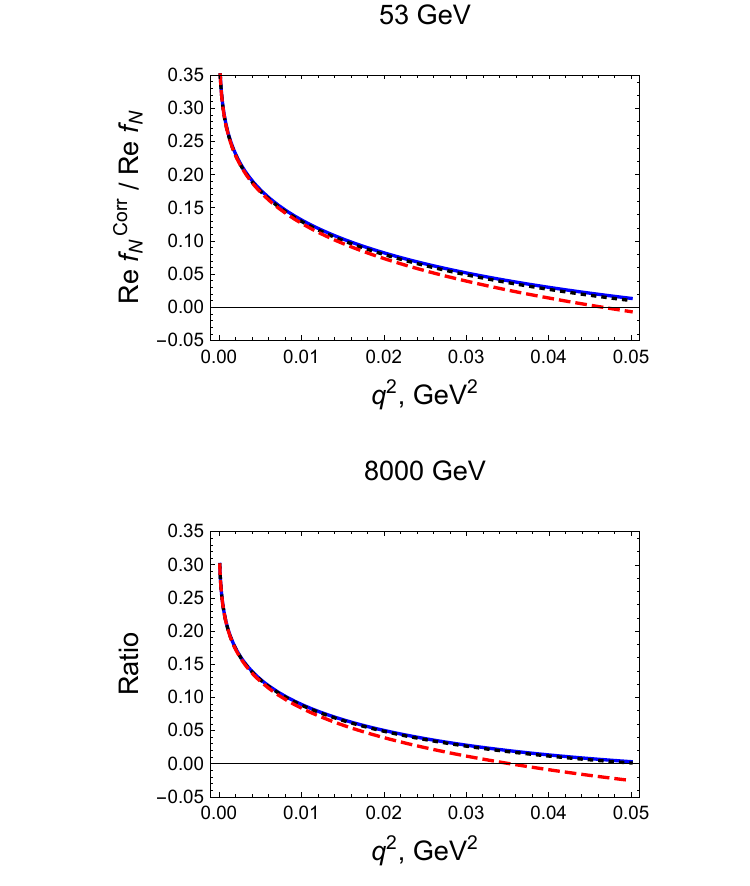}
 \caption{  Comparison of the ratios of the real parts of the Coulomb-nuclear corrections $f_N^{\,\rm Corr}$ for the standard and Gaussian form factors and the West-Yennie approximation, to the complete real part of the  corrected nuclear amplitude for the standard form factor at $W=53$ and 8000 GeV. Solid blue curves: the standard proton form factor $F(q^2)=1/(1+q^2/\mu^2)^2$ with $\mu^2=0.71$ GeV$^2$.  Dotted black curves:  the Gaussian form factor with the slope for $q^2\rightarrow 0$ matched to the slope of the standard form factor.  Dashed red curves: the West-Yennie approximation for the measured values of $B$.}
 \label{fig_WYcomp}
\end{figure}

%%%%%%%%%%%%%%%%%%%%%

In \fig{fig_WYcomp} we compare the results obtained for $\Re f_N^{\rm Corr}$ at $W=53$ with $\rho=0.8$ \cite{Amos_ISR} and 8000 GeV with $\rho=0.1$ \cite{TOTEM2016}  for three cases: (1), using the the eikonal phase in \eq{delta_FF} corresponding to the standard proton charge form factor $F_Q(q^2)=(1+q^2/\mu^2)^{-2}$ (solid blue curve); (2), using the phase  in \eq{E1} for a Gaussian form factor $f_Q^{\rm Gauss}(q^2)=e^{-2q^2/\mu^2}$ (dotted black curve); and (3), using the West-Yennie approximation $f_{N,\,WY}^{\,\rm Corr}$ in \eq{fCorr_WY} from their diagrammatic analysis (dashed red curve). In all cases we plot the ratio of the correction to the complete real part of the  corrected nuclear amplitude for the standard form factor, the quantity which appears in the Coulomb-nuclear interference. 

The results obtained using the Gaussian form factor treated with the correct eikonal phase  agree very well with those for the standard form factor through the region in $q^2$ most important for the analysis of the Coulomb-nuclear interference. This would be expected given the small-$q^2$ matching condition.  

The  West-Yennie approximation (dashed red curve) is quite good at small $q^2$ but ignores the contribution of the series in $q^2$ in \eq{Gauss_corr4}, and begins to deviate significantly from the correct result as $q^2$ increases. However,  its contribution to the complete real part---the denominator in the ratio in \fig{fig_WYcomp}---is sufficiently small  at the upper end of the $q^2$ range shown that the effect of the errors on the Coulomb-nuclear interference it likely to be minimal. That range in $q^2$  covers the 53 GeV ISR data of Amos {\em et al.}, \cite{Amos_ISR}. The errors become significant at higher $q^2$, and one should use the complete expression in \eq{Gauss_corr6} or \eq{fCorr_exp}.

The results are similar at  8000 GeV, but with increased errors in the West-Yennie approximation with increasing $q^2$. However, the region of maximum sensitivity to the Coulomb-nuclear interference simultaneously shifts to smaller $q^2$ as in \fig{Fig3} so the effect of the errors is again likely to be minimal.

%%%%%%%%%%%%%%%%%%%%%%
%%%%%%%%%%%%%%%%%%%%%%

\subsection{An application to ISR data \label{subsec:fits}}  

%%%%%%%%%%%%%%%%%%

We consider as an application of our method a reanalysis of the data on $pp$ and $\pbar p$ cross sections obtained by Amos {et al.} \cite{Amos_ISR} at the CERN ISR over the center-of-mass energy range $W= 23-62$ GeV.  The data were analyzed by those authors using the West-Yennie approximation for the Coulomb-nuclear corrections and the full range of the data assuming a purely exponential nuclear amplitude. 

The best data are for $pp$ scattering are at $W= 52.8$ GeV.  Those extend from $q^2=0.00107$ to 0.0556 GeV$^2$ with the larger values outside the region in which the simple exponential model for $f_N(s,q^2)$ is expected to hold. In particular, curvature corrections to the dominant exponential behavior  in \eq{fN_curvature} are expected to become significant for $q^2\gtrsim 0.03-0.04$ GeV$^2$ for the value $C\approx 9.8$ GeV$^{-4}$ found in the eikonal model \cite{eikonal2015},\cite{BDHHCurvature,Ha_Re_f}. The West-Yennie approximation also fails beyond that region as seen in \fig{fig_WYcomp}.

\begin{figure}[htbp]
\includegraphics{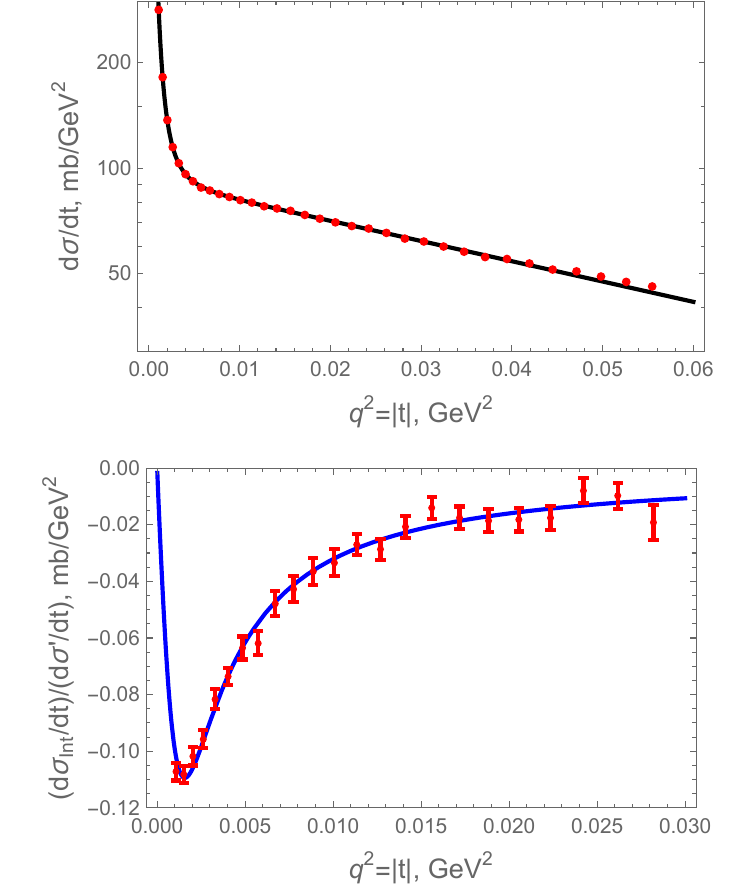}
 \caption{ The top panel shows the fit to the data on the differential cross section $d\sigma/dq^2$ for $pp$ scattering at $W=52.8$ GeV obtained in the ISR experiment of Amos {\em et al.}, \cite{Amos_ISR}. The input to the fit is described in the text. While the fit used only the data in the expected purely exponential region $q^2<0.03$ GeV$^{-2}$ for the nuclear scattering amplitude, the result is  shown over the entire $q^2$ range in which there are data. The bottom panel shows the ratio of the interference term in the differential cross section to the cross section $d\sigma'/dq^2$ with the interference term removed, over the $q^2$ range of the fit.  }
 \label{fit_53_GeV}
\end{figure}

We have restricted the data used in our analysis to $q^2<0.03$ GeV$^2$. The fit, shown in the upper panel in \fig{fit_53_GeV}, has a $\chi^2$ per degree of freedom of 0.85 compared to the value 1.46 for the fit over the entire $q^2$ range in \cite{Amos_ISR}. The accuracy of the fit at small $q^2$ is evident in the figure, as are small deviations from purely exponential behavior of the measured cross section for larger $q^2$.

%%%%%%%%%%% figure6   %%%%%%%

%%%%%%%%%%%%%%%%%%

The fitted values of the parameters are $\sigma_{\rm tot}=42.61\pm 0.07$ mb, equivalent to $A=238.2\pm 0.78$  GeV$^{-4}$, $B= 13.41\pm 0.18$ GeV$^{-2}$, and $\rho=.085\pm 0.003$.  The new values of $\sigma_{\rm tot}$ and $B$ differ from the values in \cite{Amos_ISR} 
($\sigma_{\rm tot}=42.38\pm 0.009$ mb, $B=12.87\pm 0.14$ GeV$^{-2}$) by slightly more than the quoted uncertainties, while the  new value of 
$\rho$ is at the limits compared to the previous value $\rho=0.077\pm 0.009$. These changes arise mainly from the restriction of the fit to the purely exponential region and, to a much smaller extent, from the exact calculation of the Coulomb-nuclear correction using the proper phase for the form-factor contributions as in \eq{fCorr_exp} instead of the use of the West-Yennie approximation as in \cite{Amos_ISR}.  

The changes are smaller in a similar treatment of $\pbar p$ scattering at $W=52.6$ GeV with the data (which extend to $q^2=0.039$ GeV$^2$) again restricted to $q^2<0.03$ GeV$^2$ to suppress the expected curvature effects.  The restricted fit gives $\sigma_{\rm tot}=43.61\pm 0.33$ mb 
($A=249.5\pm 3.8$ GeV$^{-4}$), $B=13.71\pm 0.65$ GeV$^{-2}$, and $\rho=0.0974\pm 0.0122$.  The differences from the fit of Amos {\em et al.} over 
the entire range in $q^2$ using the West-Yennie approximation ($\sigma_{\rm tot}=42.32\pm 0.34$ mb, $A=249.5\pm 3.8$ GeV$^{-4}$,  $B=13.03\pm 0.52$ GeV$^{-2}$, $\rho=0.106\pm 0.016$) are all small and within the quoted uncertainties which are significantly larger than in $pp$ scattering.

At the remaining ISR energies with good data, $W=23.5$ and 30.6 GeV for $pp$ scattering, and $W=30.4$ GeV for $\pbar p$ scattering, the measured differential cross sections  are already restricted to the ranges $q^2<0.0102,\ 0.176$,  and 0.0156 GeV$^{-2}$, all within the expected exponential regions for the nuclear cross 
sections. The West-Yennie approximation is quite accurate in these ranges as seen in \fig{fig_WYcomp}, so there are no measurable changes in $\sigma_{\rm tot},\ B$ and $\rho$ with a change to the exact treatment of the form-factor phase.

In the bottom panel in \fig{fit_53_GeV} we show the ratio of the interference term found in differential $pp$ scattering cross section at $W=52.8$ GeV to the cross section with that term removed over the range of the fit, $0.00107\, {\rm GeV}^{-2}\leq q^2\leq 0.0282\,  {\rm GeV}^{-2}$. The experimental and theoretical results agrees remarkably well as was indicated by the $\chi^2$ per degree of freedom for the fit.  The corresponding figure for $\pbar p$ scattering at 52.6 GeV (not shown) again displays good agreement of the theoretical results for $d\sigma/dq^2$ and the interference ratio with the data of Amos {\em et al.}, but the experimental uncertainties are much larger as noted above.  See for reference Fig.\ 7 in \cite{Amos_ISR} which also shows the interference ratios at the remaining energies.

We have not re-examined the results obtained by the TOTEM Collaboration at 8,000 \cite{TOTEM2016} and 13,000 \cite{TOTEM2019_Rho_13} GeV. Those authors included calculations of the Coulomb-nuclear corrections using the much more complicated method of Kundr\'{a}t and M. Lokaji\v{c}ek \cite{KL-Coulomb}, not shown in the papers, and fit  the curvature corrections over the comparatively much wider range of $q^2$ covered by the TOTEM data with results that agreed reasonably well with those calculated in the comprehensive eikonal model in \cite{BDHHCurvature,eikonal2015,eikonal_update}.

%%%%%%%%%%%%%%%%%%%%%%
%%%%%%%%%%%%%%%%%%%%%%

\section{Conclusions \label{sec:conclusions}}
 
We have presented a very simple method for the calculation of the mixed Coulomb-nuclear corrections to the $pp$ and $\pbar p$ scattering amplitudes  through \eq{corr_int2} or \eq{double_int} which is applicable for any reasonable model for $f_N$.  The sum of the analytic expression for the pure Coulomb and form-factor terms in \eq{f_Coul}, the model nuclear amplitude $f_N$, and this small correction term gives our expression for the full $pp$ or $\pbar p$ scattering amplitude $f(s,q^2)$, \eq{f_defined}. This approach provides a substantial improvement  in clarity and simplicity relative to the methods most commonly used at present \cite{Cahn,KL-Coulomb,Petrov1}.  

We showed that the results obtained using the simple exponential model for $f_N(s,q^2)=(i+\rho)\sqrt{A/\pi}e^{-\frac{1}{2}Bq^2}$ commonly used to fit data on the differential cross section $d\sigma/dq^2$ at high energies  agree in the relevant $q^2$ ranges with those obtained in the comprehensive eikonal model of Block {\em et al.} \cite{eikonal2015,eikonal_update} at energies from $W=52.8$ GeV to 13 TeV.   We then used the model to explore the expected interference effects in $pp$ and $\pbar p$ scattering, possible effects of the Martin zero in the real part of the scattering amplitude on the determination of $\rho$, and the corrections to the West-Yennie approximation for the Coulomb-nuclear correction, and concluded with a reanalysis of the ISR data of Amos {\em et al.} \cite {Amos_ISR}.

We emphasize that the method is not confined to the exponential type models for the nuclear amplitude at high energies, but can be applied to Regge-type and other models as well, and can also be used for proton-nucleus scattering.

%\bibliography{CoulombMathbib}

\begin{thebibliography}{99}
\bibitem{corr1} The expression given for $2\delta_{FF}$ in Eq.~(17) of \cite{DH_CoulNucl} is missing the factor 2 and has the wrong sign and summation limits on the right-hand side as given there.
\bibitem{CKLapproach} It is at this point (but with the last two terms in \eq{f_defined} combined) that the complications of the Cahn approach enter.  Cahn \cite{Cahn} wrote the first factor in the integrand in \eq{corr_int2} as the inverse Fourier-Bessel transform of the combined Coulomb and form factor amplitude, and then used the convolution theorem to write the product of the two transforms in the final integral as a two-dimensional convolution of the original amplitudes.  This is not strictly allowed. The integral which defines the transform of the first Coulomb-related factor in \eq{corr_int2} is divergent so must be regularized in some way.  Cahn circumvented this problem by rearranging the final integrals  to obtain a convergent result, also introducing some further approximations. The same approach was later followed by  Kundr\'{a}t and M. Lokaji\v{c}ek \cite{KL-Coulomb}. See also \cite{Nekrasov}. This manipulation is not necessary: the phases in the first factor in the integrand in  \eq{corr_int2} are known and the integral converges rapidly as it stands.
\bibitem{analytic_evaluation} 
To evaluate the first terms in the integral in \eq{fCorr_exp}, note that 
\begin{eqnarray*}
\frac{1}{B}\int_0^\infty db b\left[\ln{\left(\frac{qb}{2}\right)}+\gamma\right]e^{-b^2/2B}J_0(qb)&=&\frac{1}{2B}\frac{d}{dc}\int_0^\infty db b\left[\frac{d}{dc}\left(\frac{qb}{2}\right)^{2c}+2\gamma\right] e^{-b^2/2B} J_0(qb) \Bigg\vert_{c=0} \\
&=& \frac{1}{2}\frac{d}{dc}\left[\left(\frac{q^2B}{2}\right)^c\Gamma(1+c)\, _1F_1(-c;1;q^2B/2)+2\gamma\right]_{c=0}e^{-Bq^2/2} \\
&=& \frac{1}{2}e^{-Bq^2/2}\Big(\ln{\frac{q^2B}{2}}+\gamma-\sum_{k=1}^\infty \frac{1}{kk!}(q^2B/2)^k\Big)
\end{eqnarray*}
where the integral in the first line follows from \cite{Watson}, Eq.~13.3(3), or \cite{dlmf} 10.22.54. This result agrees with that obtained by West and Yennie \cite{WestYennie} by a quite different method in their diagrammatic analysis.
\vspace*{1ex}

The remaining terms in \eq{fCorr_exp} associated with the form factors  involve the integrals 
\begin{eqnarray*}
\int_0^\infty db b (\mu b)^mK_m(\mu b)e^{-b^2/2B}J_0(qb)=\int_0^\infty db b (\mu b)^mK_m(\mu b)e^{-b^2/2B}\left(1-\frac{(qb)^2}{4}+\frac{(qb)^4}{64}-\cdots\right).
\end{eqnarray*}
The series can be truncated or ignored for small $q^2$ and the remaining integrals evaluated analytically in terms of Whittaker functions $W_{\mu,\nu}(z)$ (\cite{dlmf}, Sec.~13.14, \cite{HTF}, Sec.~6.9) using \cite{TIT}, 10.2(25). These are related to confluent hypergeometric functions of the second kind  and are not elementary or simple to use. By far the simplest approach to the calculation of the integral and that adopted here  is direct numerical evaluation.
\bibitem{Phi_definition} In Bethe's summary of his calculations in \cite{BetheCoulomb}, he put the inverse phase on the Coulomb term, allowed if the amplitude factors as assumed. Since it does not, we keep the correction on the nuclear amplitude where it originates, and will expand the exponential to obtain what is in effect Bethe's approximation for $f_N^{\rm Corr}$,  $f_N^{\rm Corr}\approx-i\Phi f_N(s,q^2)$.
\bibitem{factorization} Although the complete amplitude does not factor in general, we can factor out the $q^2$-dependent part of the Coulomb phase in \eq{nucl_combined}. This term could potentially cause problems with the ``Bethe approximation" just discussed because of the divergence of $ln{(q^2)}$ for $q^2\ll 1$, the region of maximum sensitivity to the interference.  This is especially important for proton-nucleus scattering, Bethe's original problem, where the nuclear charge $Z$ can be large enough to make the corrections non-negligible. To see how this can be handled, write the Coulomb phase as
%
\bdisplay  
2\delta'=2\eta\left(\ln{(qb/2)}+\gamma\right)=\eta\left(\ln{(q^2a^2)}+\gamma\right)+\eta\left( \ln{(b^2/4a^2)}+\gamma\right)
\edisplay
%
 with $a$ constant. The factor in $e^{2i\delta_C'}$ corresponding to the first part of this expression is constant in $b$ and factors out of the integral. The remaining $b$-dependent term does not, but the integral involving that term vanishes to first order in $\eta$ at $q^2=0$ in the exponential model for ${\widehat f}_N$ for $a=\sqrt{B/2}$. The remainng corrections associated with the form factors are sufficiently small that they can reasonably be exponentiated, with $1-i\Phi\rightarrow e^{-i\Phi}$, giving Bethe's result with nonsingular corrections of at most $O(\eta^2)$. Similar arguments should hold for other forms of the nuclear amplitude.
\end{thebibliography}

\end{document}